\documentclass[11pt,titlepage]{article}

\usepackage{setspace} 

\usepackage{amsmath}

\usepackage{graphicx}

\usepackage[round,numbers,sort&compress]{natbib} 
\title{Thermodynamic pathways to genome spatial organization in the cell nucleus}

\author{Mario Nicodemi\thanks{
           Corresponding author.  Address: 
           Department of Physics,
	   University of Warwick, UK
	   } \\
	Department of Physics and Complexity Science,\\ 
	University of Warwick, UK \& INFN Napoli, Italy
	\and Antonella Prisco \\
	CNR Inst. Genet. and Biophys. `Buzzati Traverso',\\ 
	Via P. Castellino 111, Napoli, Italy}

\date{}

\begin{document}

\maketitle

\abstract{
The architecture of the eukaryotic genome is characterized by a high degree of spatial organization. Chromosomes occupy preferred territories correlated to their state of activity and, yet, displace their genes to interact with remote sites in complex patterns requiring the orchestration of a huge number of DNA loci and molecular regulators. Far from random, this organization serves crucial functional purposes, but its governing principles remain elusive. 
By computer simulations of a Statistical Mechanics model, we show how architectural patterns spontaneously arise from the physical interaction between soluble binding molecules and chromosomes via collective thermodynamics mechanisms. 
Chromosomes colocalize, loops and territories form and find their relative positions as stable thermodynamic states. These are selected by ``thermodynamic switches" which are 
regulated by 
concentrations/affinity of soluble mediators and by number/location of their attachment sites along chromosomes. 
Our ``thermodynamic switch model" of nuclear architecture, thus, explains on quantitative grounds how well known cell strategies of upregulation of DNA binding proteins or modification of chromatin structure can dynamically shape the organization of the nucleus. 
}

\clearpage

\section*{Introduction}

Within the cell nucleus, genome structure has a complex organization in space spanning different scales. Chromosomes tend to form a set of distinct territories and, at a smaller level, are folded in higher-order structures, while a variety of physical intra- and inter-chromosomal interactions between specific DNA sequences has been reported  \citep{Cremer2006,Lanctot2007,Misteli2007,Meaburn2007,Fraser2007,Gasser07}.
While structures can be formed by tethering specific DNA segments to scaffolding elements, such as the nuclear envelope, DNA-DNA contacts and chromatin loops are an ubiquitous organizational feature 
extending up to hundreds of kilobases, and relocating, for instance, genes to substantial distances outside of their territory. 
Intriguingly, relative positions of territories, as well as of DNA sequences within a territory, have a probabilistic nature dynamically changing with cell type and cell cycle phase. Yet, stable, non-random patterns are established, fundamental to genome regulation, as disruptions relate to serious diseases, most notably, cancer \citep{Cremer2006,Lanctot2007,Misteli2007,Meaburn2007,Fraser2007,Gasser07}. 
Remarkably common features are shared in chromatin organization processes, 
but the underlying principles of their control in space and time are still largely mysterious \citep{Misteli2007}. 

While there is evidence that far apart DNA sequences, even on different chromosomes, can come together by interacting with molecular factors, the mechanisms whereby they do so and higher-order structures and territories arise are still largely mysterious. 
One of the scenarios proposed to explain the establishment of contacts between DNA elements is the so called `random collision' picture (see, e.g., \citep{deLaat03})  
whereby chromatin flexibility allows factors bound to one sequence to randomly contact factors bound to surrounding chromatin. 
Although active mechanisms of directed motion have been described (see, e.g., \citep{Chuang2006}), 
diffusion-based mobility is indeed a prevailing mechanisms that delivers molecular complexes to their specific nuclear targets (see \citep{Misteli2001} and ref.s therein). So, loops could be formed when a diffusing factor succeeds in bridging two chromosomal sites as a result of a ``random double encounter", whereby the molecule by chance encounters its first binding site and then, by chance, the second one. 
Yet how such loops persist beyond the initial `random collision' is totally unclear 
\citep{deLaat03,deLaat07} and many questions remain open:
how strong are the bonds required to hold in place whole chromosomal segments? 
How are stochastic encounters coordinated in space and time for a functional purpose by the cell? 
Can higher-order structures and territories spontaneously arise from them?
Here, by use of a polymer physics model we propose a scenario to answer such questions.

Sequence-specific DNA-binding molecular factors have emerged as critical regulators of chromatin interactions in the nucleus 
\citep{Cremer2006,Lanctot2007,Misteli2007,Meaburn2007,Fraser2007,Gasser07} and some of them are encountered in a variety of cases, as for instance SATB1 \citep{Galande}, Ikaros \citep{Brown1997}, PcG \citep{Bantignies2003}, and CTCF Zn-finger proteins, the latter known to mediate also interchromosomal contacts \citep{Kurukuti2006,Ling2006,Zhao2006,Lee07}.
In some cases a combination of factors is required to induce looping, as in the example of the erythroid transcription factor GATA-1 and its cofactors 
at the $\beta$-globin locus \citep{Drissen04,Vakoc05}. 
Analogously, GATA-3 and STAT6 cooperation has been proposed to establish long-range chromatin interactions at the TH2 cytokine locus \citep{Spilianakis04}. 
Transcription factories themselves, i.e., local high concentrations of Pol II, have been proposed to act as hubs in the formation of loops and the colocalization of distant genes, even outside chromosome territories (see \citep{Osborne2004,Fraser2007,Marenduzzo07} and ref.s therein). 
In the last few years, protein-DNA interactions that occur in vivo have been probed by innovative genome-wide techniques 
leading to the description of thousands of binding sites for 
DNA binding proteins 
\citep{Massie2008}, 
and systematic approaches to measuring their binding energy landscapes are being developed \citep{Maerkl2007}.
DNA binding proteins typically exhibit a number of target loci, which can be found  clustered in groups. Their DNA chemical affinities are in general found in the weak biochemical energy range, $E_X\sim 1\div 15 kT$ \citep{Maerkl2007,Siggia05,Hwa02,Lassig07,Berg08} ($k$ is the Boltzmann constant and $T$ room temperature). Although in most cases only qualitative information is available, details on binding energies and DNA locations have been clearly described for a number of examples (see \citep{Maerkl2007,Siggia05,Hwa02,Lassig07,Berg08} and Ref.s therein). Initial works on bacteria have shown that DNA binding proteins can have hundreds of DNA sites with affinities in the range $2\div15kT$ (\citep{Hwa02} and ref.s therein). In yeast, more recently, the landscape of binding energies and loci has been explored by advanced computational biophysics methods: the distribution of their binding energies spans a range of about $10kT$, and they can have hundreds of DNA binding sites across the genome as well (see \citep{Berg08} and ref.s therein). Similar ranges in binding energies have been found in higher eukaryotes, including mice and humans (\citep{Siggia05,Maerkl2007,Massie2008} and ref.s therein), where common examples exist of proteins with thousands of DNA target sequences.

DNA-DNA interactions mediated by molecular factors are being extensively mapped, revealing a complex network of intra and interchromosomal interactions \citep{Simonis2007}. 
Clusters of binding sites of SATB1 \citep{Galande,Purbey08}, and zinc finger class proteins CTCF \citep{Donohoe07,Kurukuti06}, Ikaros \citep{Brown1997} and GATA-1 \citep{Jing08} were found in a number of regions involved in DNA cross talk. 
An important example is the cluster of CTCF binding sites responsible for X chromosome pairing, at the onset of X Inactivation, located at the {\em Xist/Tsix} locus where, in a few kb short sequence, a group of about hundred binding sites, each 20b long, is found  \cite{Lee07,Donohoe07}.
Expansion of the nuclear volume leads to the disassembly of several nuclear compartments 
\citep{Hancock2004} which might suggest that specific 
concentrations of macromolecules are required for the self-assembly of nuclear structures. 
Loss of specific interchromosomal DNA-DNA contacts has been described after a marked reduction, for instance, of the amount of CTCF \citep{Ling2006,Lee07}. 
Changes in the concentration of ``heterochromatin" proteins, e.g., HP1  \citep{Gasser06,Workman08}, are also known to affect the organization of genomic DNA \citep{Locke1988}.

The conformational properties of chromosomes have been investigated by using polymer models in the past \citep{Hahnfeldt93,Yokota95,Sachs95,Munkel98,CremerLangowski,CremerCremer,Schiessel,MarenduzzoMicheletti,MarenduzzoCook,Bohn07,nicodemi07a,nicodemi07b}. The chromatin fiber was modeled as a random walk in a confining geometry \citep{Hahnfeldt93}, and the possibility 
was considered to include giant loops, of about 3 Mb, departing 
from its backbone to describe folding at different 
scales (RWGL model \citep{Yokota95,Sachs95}). The multi-loop-subcompartment MLS model \citep{Munkel98,CremerLangowski} aimed to represent `rosette' structures, with 120kb loops, like those experimentally observed. 
To describe the radial arrangement of chromosome territories in human cell nuclei, a model was proposed \citep{CremerCremer} where each chromosome is approximated by a linear chain of spherical 1 Mbp-sized chromatin domains. Adjacent domains are linked together by an entropic spring and by an effective excluded volume potential, while to maintain the compactness of chromosome territories a weak potential barrier around each chromosome chain was also included. 
Recently, the ``Random Loop" polymer model \citep{Bohn07} has introduced the idea that a set of randomly located sites along a random walk chain can bind each other, in order to explain, at the same time, the experimentally observed presence of loops of different scales and the leveling-off of the mean square distance between two beads of the chain at genomic distances above 1-2Mb. 
Several other chromatin features have been successfully explored by computer simulations, including nucleosome interactions \citep{Schiessel}, packing \citep{MarenduzzoMicheletti,MarenduzzoCook}, molecular assembly \citep{nicodemi07a,nicodemi07b}, providing a vivid description of the geometry and conformational properties of chromatin as observed in experiments. 

Here, by investigation of a polymer physics model inspired by the above biological scenario, we discuss how architectural patterns spontaneously arise from the interaction of soluble binding molecules and chromosomes. 
Our model shows that thermodynamics dictates pathways to complex pattern formation: loops, colocalization of distant sequences, chromosomal domains, structures and territories spontaneously organize as stable thermodynamic states when specific threshold values in molecule concentrations or their affinity to DNA sites are exceeded.
By regulation of expression levels and modification of DNA targets, the cell can, thus, act on ``thermodynamic switches" \citep{npp1,npp2} to reliably control its genome organization in space and time.

\section*{Theoretical Model}

To describe a system made of a chromosome and its binding molecules, we consider an established model of polymer physics \citep{EdwardsDoi,BinderHeermann}: the chromosome polymer is modelled as a Brownian self-avoiding walk (SAW) of $n$ non-overlapping beads, and soluble molecules as Brownian particles having a concentration, $c$ (see Fig.\ref{equil_picts}). 
A fraction, $f$, of polymer sites can bind the diffusing molecules, with a chemical affinity $E_X$ in the weak biochemical range (see Methods for details). Here, for sake of simplicity, binding sites are uniformly interspersed with non-binding regions along the chain. Each molecular factor can simultaneously bind many a site on the polymer, a feature that reflects the presence of multiple DNA binding domains in a number of regulatory proteins (e.g., CTCF). Mediating molecules with only one DNA binding site, that are able to interact with each other, could be also considered; since a group of linked molecules can be represented, in the model, as just one mediator, the picture is unchanged. 
The equilibrium thermodynamic properties of such a system were determined by extensive Monte Carlo simulations \citep{BinderHeermann,Binder}. 

\subsection*{Methods}

In our Monte Carlo computer simulations \citep{BinderHeermann} molecules and polymers diffuse in a cubic lattice having a linear size $L$, and its spacing, $d_0$, sets the space unit. For computational purposes, we mostly consider lattices of linear size $L=32$, though, we tested our results up to $L=128$. 
SAW polymer beads have a diameter, $d_0$, and each bead in a chain is on a next or nearest next neighboring site of its predecessor. Molecules (of size $d_0$) are also subject to Brownian motion. When neighboring a binding site of a polymer, molecules interact with it via an effective energy, $E_X$. According to the studied case (see Results), up to six distinct sites (i.e., the nearest neighbors in a cubic lattice) on the same chromosome, or alternatively two sites on different chromosomes can be bound at the same time. 

Our schematic model is a coarse-grained description of a real polymer and, since by now we mostly focus on the description of a general conceptual framework, beads only represent generic binding sites (they could be a binding locus, the bases of specific binding sequences, etc.). In cases where detailed data on binding sequences and regulator chemistry is available, such information could be easily taken into account in the model to produce specific quantitative predictions. The role of interactions with, e.g., the nuclear membrane could be also included, but to make the message simpler, we decided not to discuss such an aspect here. 

To obtain thermodynamic equilibrium configurations, the Metropolis Monte Carlo method was applied. Chromosome polymers are initially equilibrated in a random self-avoiding configuration obtained, in absence of binding molecules, by random displacements of single beads under the constraint that each bead in the chain is on a next or nearest next neighboring site of its predecessor. Then molecules are inserted at random empty positions in the lattice to attain a given concentration. In the ensuing Metropolis Monte Carlo procedure, a sequence of states is generated by a Markov process \citep{BinderHeermann} whereby a new position for a particle/bead is stochastically selected according to a specific transition matrix satisfying the `principle of detailed balance' which in turn guarantees the convergence in probability of the sampled states to Boltzmann thermodynamic equilibrium distribution. 
The transition probability for a particle/bead to diffuse to a neighboring empty site is proportional to the Arrhenius factor $r_0\exp(-\Delta E/kT)$, where $\Delta E$ is the energy barrier in the move, $k$ the Boltzmann constant and $T$ the temperature \citep{BinderHeermann}. The lattice has periodic boundary conditions to reduce boundary effects. 

In a Monte Carlo lattice sweep every particle and bead in the system, randomly selected, is updated on average once. Our simulations run for up to $10^9$ Monte Carlo lattice sweeps as the number of decorrelation steps from an initial configuration can be as large as $10^5$. 
The achievement of stationarity was monitored by checking the dynamics 
of different quantities, such as the system gyration radius, the distance between two polymers, the system energy and the number of particles attached to polymers. Once equilibrium is reached for all these quantities, thermodynamic averages are calculated by considering only configurations having a distance larger than the decorrelation length. Finally, averages are also performed over up to 2048 runs from different initial configurations. 
Confidence intervals are calculated as squared deviations around these averages, as discussed in \citep{BinderHeermann}; they are indicated in our figures by the size of the used symbols.

Our code has two core routines, well described in Binder and Heermann \citep{BinderHeermann}: the ``lattice gas" spin-exchange Metropolis routine for particle displacement, and the Self Avoiding Walk routine. Several means were considered to avoid algorithmic errors, as those suggested in \citep{BinderHeermann}. 
Each different routine in the code was tested independently. For example, the routine generating the evolution of the Self Avoiding Walk chain was tested by checking the behavior of the calculated average gyration radius, $R_g$, against the chain length, $n$, and the power law $R_g \sim n^{\nu}$ with an exponent $\nu\sim 0.6$, well established in the literature \citep{EdwardsDoi,BinderHeermann}, was recovered. An other internal test was to show that other geometric quantities, such as the chain end-to-end distance did scale in the same way as $R_g$.

Real chromosomes differ in size (i.e., $n$) and arrangement of their binding sites. Such differences affect their specific behaviors, but the general picture we aim to depict here is not altered by changes to the selected values of these parameters (e.g., $n$ and $L$). To make computation time feasible, we mostly use $n=64$, but we tried $n$ as large as $128$. 
The robustness of our model is well established in polymer physics \citep{EdwardsDoi,BinderHeermann}, and to check the effects of finite size scaling we explored changes of the polymer chain length in the range $n\in\{16,32,64,128\}$ (see Results). 

\section*{Results}

\subsection*{Intrachromosomal interactions, loop and territory formation}

\begin{figure}
   \begin{center}
      \includegraphics*[width=5in]{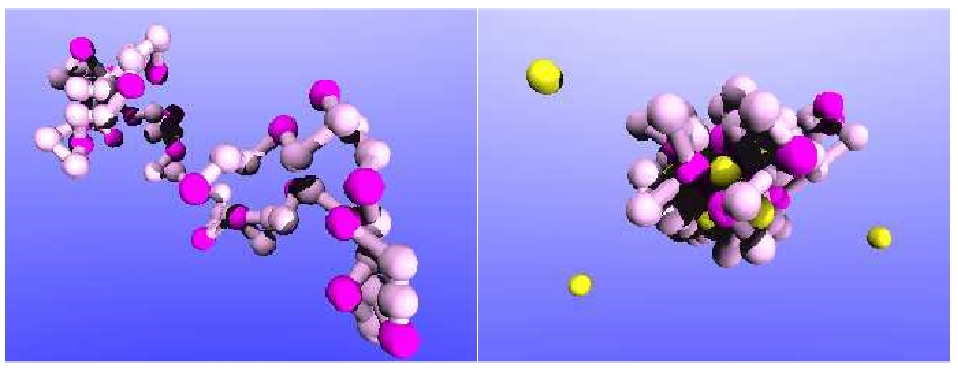}
      \caption{}
      \label{equil_picts}
   \end{center}
\end{figure}

We first discuss how a chromosome can fold up in loops within a territory with a specific spatial conformation by interacting with soluble molecules, and how the process can be controlled by the cell (see Fig.\ref{equil_picts}). The folding state of the polymer is illustrated by its squared radius of gyration, $R_g^2$ \citep{EdwardsDoi}:
$R_g^2={1/(2n^2{\cal N})}\sum_{i,j=1}^{n} ({\bf r}_i-{\bf r}_j)^2$, 
where ${\bf r}_i$ is the position of bead $i\in\{1,...,n\}$, and ${\cal N}$ a 
normalization constant 
(here $\cal N$ equals the average squared gyration radius of a randomly floating SAW chain of size $n$). 
$R_g$ represents the radius of a `minimal' sphere enclosing the polymer: it attains a maximum when the polymer is loose and randomly folded, and a minimum when loops enclose it in a compact lump.

\begin{figure}
   \begin{center}
      \includegraphics*[width=2.25in,angle=-90]{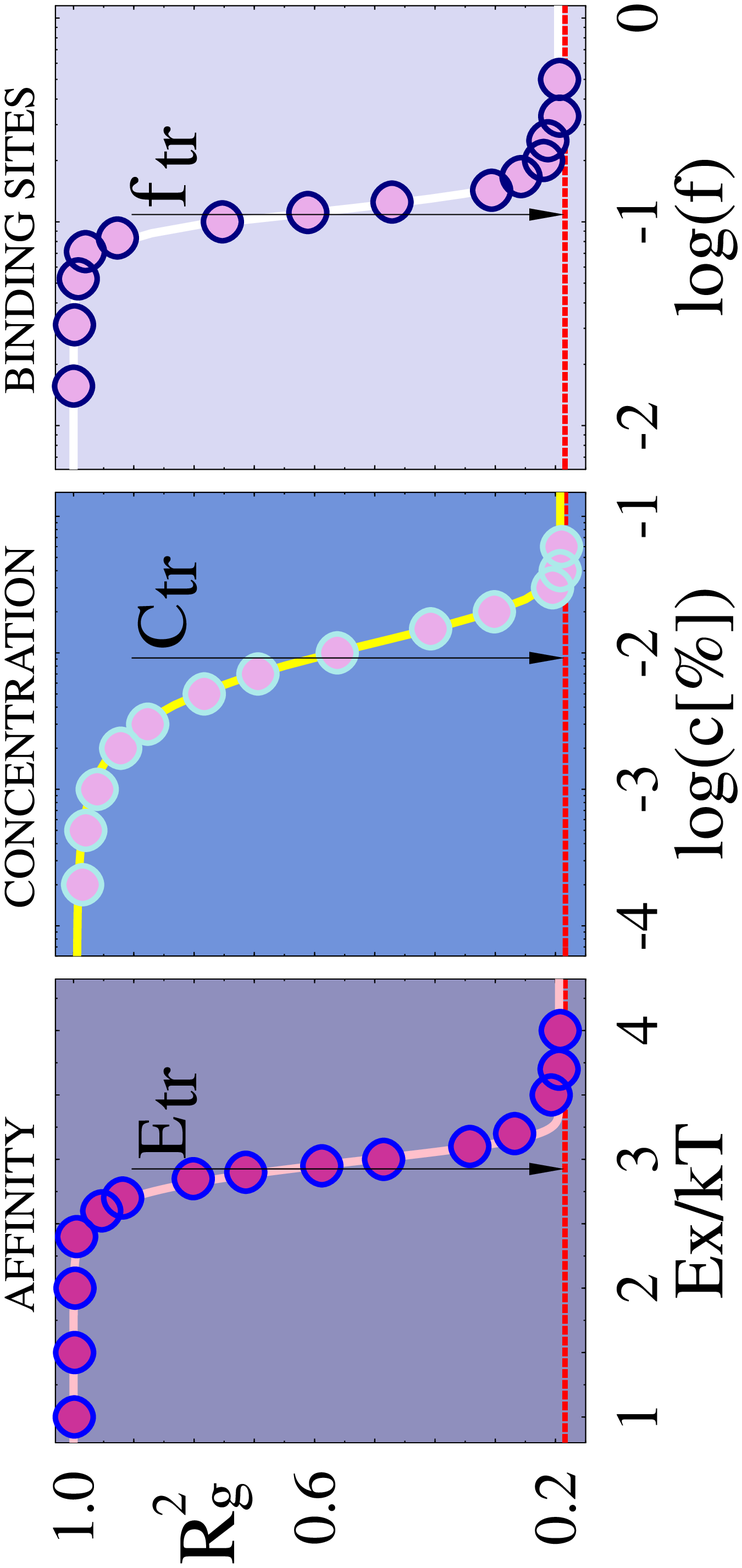}
      \caption{}
      \label{equil_ex}
   \end{center}
\end{figure}

In presence of a given concentration of molecules, loops could be created by chance when a particle bridges a couple of chromosomal sites having a non zero affinity, $E_X$. Fig.\ref{equil_ex} left panel shows, indeed, that $R_g^2$ attains a small plateau value when $E_X$ is large enough (say above the inflection point, $E_{tr}$, of the curve $R_g^2(E_X)$): bridges are thermodynamically favored and the polymer takes a compact looped territorial conformation, as seen in a typical `snapshot' from computer simulations depicted in Fig.\ref{equil_picts} right panel. The system behavior, however, switches for $E_X < E_{tr}$, since $R^2_g$ keeps its maximal value corresponding to a fully open polymer floating in space (see Fig.\ref{equil_picts} left panel) and no stable loops are formed. The folding level also depends on factors such as concentration of molecules, number and location of DNA binding sites (see below).

The above results have an intuitive basis: if $E_X$ is small the half-life of a randomly formed bridge is small and polymer segments on average float away; the higher $E_X$, the higher the number of bound molecules and, thus, of bridges which reinforce each other and stabilize the conformation, as multiple bonds should be simultaneously broken to release a loop. Our physics model reveals, in particular, that a precise threshold marks the switch between the two regimes; $E_{tr}$ corresponds to a thermodynamic phase transition \citep{Stanley}, as discussed later on. 
This picture illustrates on quantitative grounds how chromatin modifications, such as DNA methylation or post-translational modifications of DNA binding proteins (well described cell strategies to change genomic architecture), can result in dramatic, switch-like,  effects. 

In a different thermodynamic pathway to loop formation, the cell can regulate the concentration, $c$, of binding molecules. The plot of $R^2_g(c)$ (Fig.\ref{equil_ex} central panel) shows how $c$ affects the compaction state 
of the polymer. When $c$ is below the threshold, $c_{tr}$, $R^2_g(c)$ has a value corresponding to random folding, while above $c_{tr}$, it decreases towards its ``looped state" value. A broad crossover region is found around $c_{tr}$, revealing that $R^2_g$, which can be envisioned in our example as the radius of the ``territory", 
can be tuned across a range of values. So, the regulation of a DNA binding protein concentration (a typical event in cellular behavior) can act as an other switch to reliable assembling of genomic architectures.

Finally, we find (Fig.\ref{equil_ex}, right panel) that a minimal threshold in the number of polymer binding sites (or in their fraction $f$) is required for stable looping/territory formation. Conceptually, the case of a polymer with a low number of binding sites is equivalent to the case of a polymer with many binding sites in the presence of a limiting concentration of mediators. The function $R^2_g(f)$ indicates that a ``thermodynamic switch" to DNA compaction resides in the potential to obliterate/restore a fraction of sites via chromatin modifications that abolish binding of the relevant regulatory molecule. Intriguingly, the presence of a thermodynamic threshold in $f$ could relate to the experimental observation that multiple binding sites for mediators have been found at chromosome interaction loci and looping points (e.g., CTCF mediated interactions). 
Importantly, in our model we find that the threshold value, $f_{th}$, is a strongly decreasing function of the binding energy, $E_X$. 
This can be expected as, for an above threshold mediator concentration, $c$, the overall binding energy linking two polymer strands is approximately $fE_X$; so an increase of $E_X$ would correspond to an inversely proportional reduction of $f_{tr}$.

\begin{figure}
   \begin{center}
      \includegraphics*[width=3.25in]{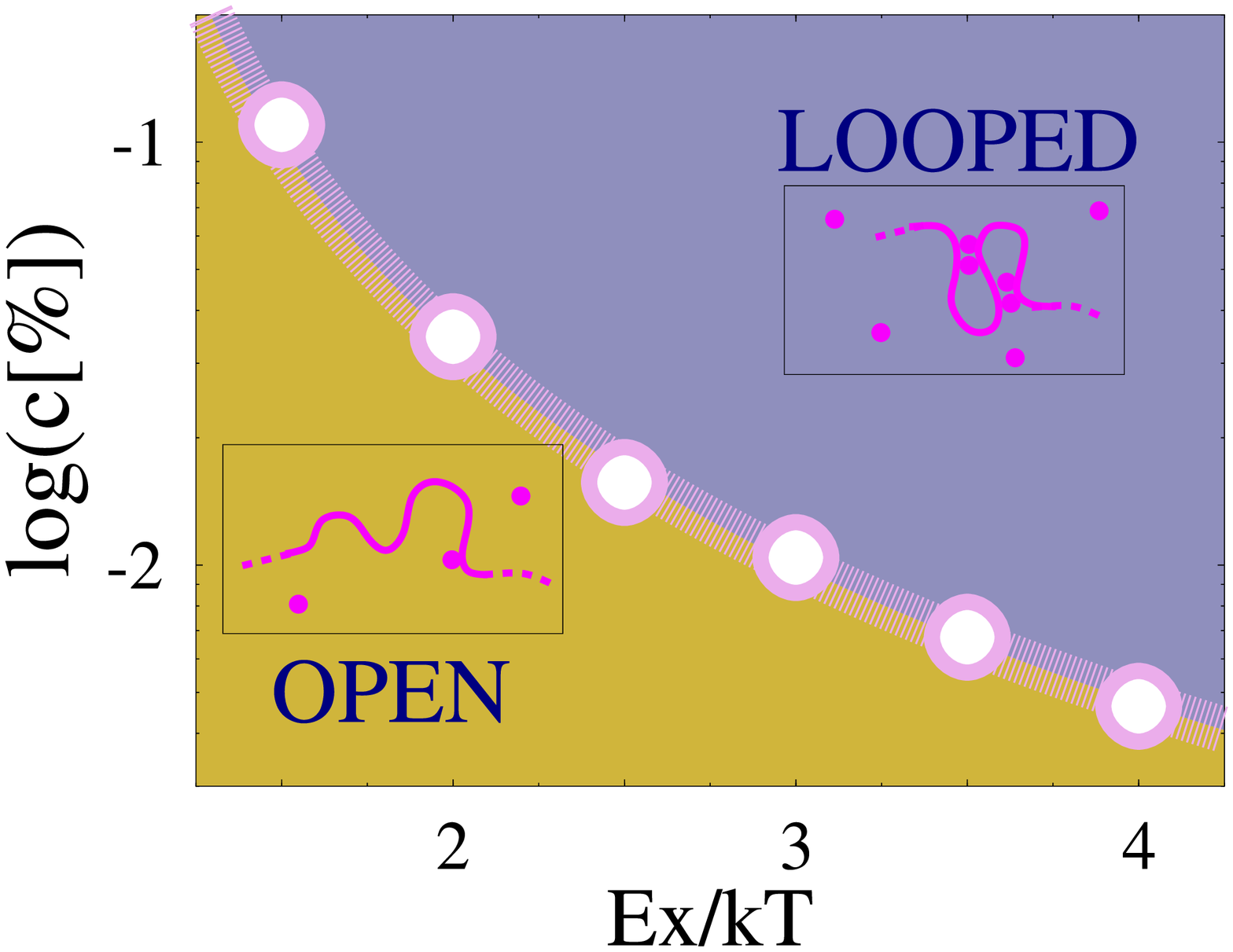}
      \caption{}
      \label{ph_diag}
   \end{center}
\end{figure}

The above described ``thermodynamic switches" define a robust regulatory mechanisms as seen in the phase diagram of Fig.\ref{ph_diag}, reporting the equilibrium state of the chromosome (open vs looped) in a wide range of $E_X$ and $c$ values (for a given $f$). In particular, Fig.\ref{ph_diag} shows that the threshold value $E_{tr}(c)$  (dashed line) required for loop assembly decreases as $c$ increases and can be as weak as an hydrogen bond.  
In the cell, the possibility to drive looping by use of sites with even low binding energy for their soluble ligands could be important to prevent polymers from getting stuck in topologically unacceptable entanglements or ectopic associations, since each single low energy bond can be easily broken for adjustments.

The threshold values in the $(c,E_X,f)$ space (see Fig.\ref{ph_diag}), related in polymer physics to the chain $\theta$-point \citep{EdwardsDoi}, correspond to a phase transition occurring in the system when one of two competing thermodynamics mechanisms prevails: entropy, $S$, which favors loose random folding, or energy, $E$, which increases when bonds between molecules and DNA sites are established by loop formation. The system spontaneously tends (as it is finite sized \citep{Stanley}) to select the state where its Free Energy, $F(c,E_X,f)=E-TS$, is minimized. 
More precisely, the chromosome conformation has a specific stochastic distribution (having a width which can be very narrow) following from Boltzmann thermodynamics weights \citep{Stanley}. 

\subsection*{Scaling behaviour of the model}

As molecule binding regions on `cross-talk' loci of real chromosomes have variable sizes, $n$, we explored the `scaling behaviour' of our system by varying the polymer chain length in the range $n\in\{16,32,64,128\}$, for the above value of the containing box size $L$. 
The reference case considered previously, and in the rest of the paper, has $n=64$, which is comparable to values included in similar studies \citep{Munkel98,CremerLangowski,CremerCremer}. 

\begin{figure}
   \begin{center}
      \includegraphics*[width=3.25in,angle=-90]{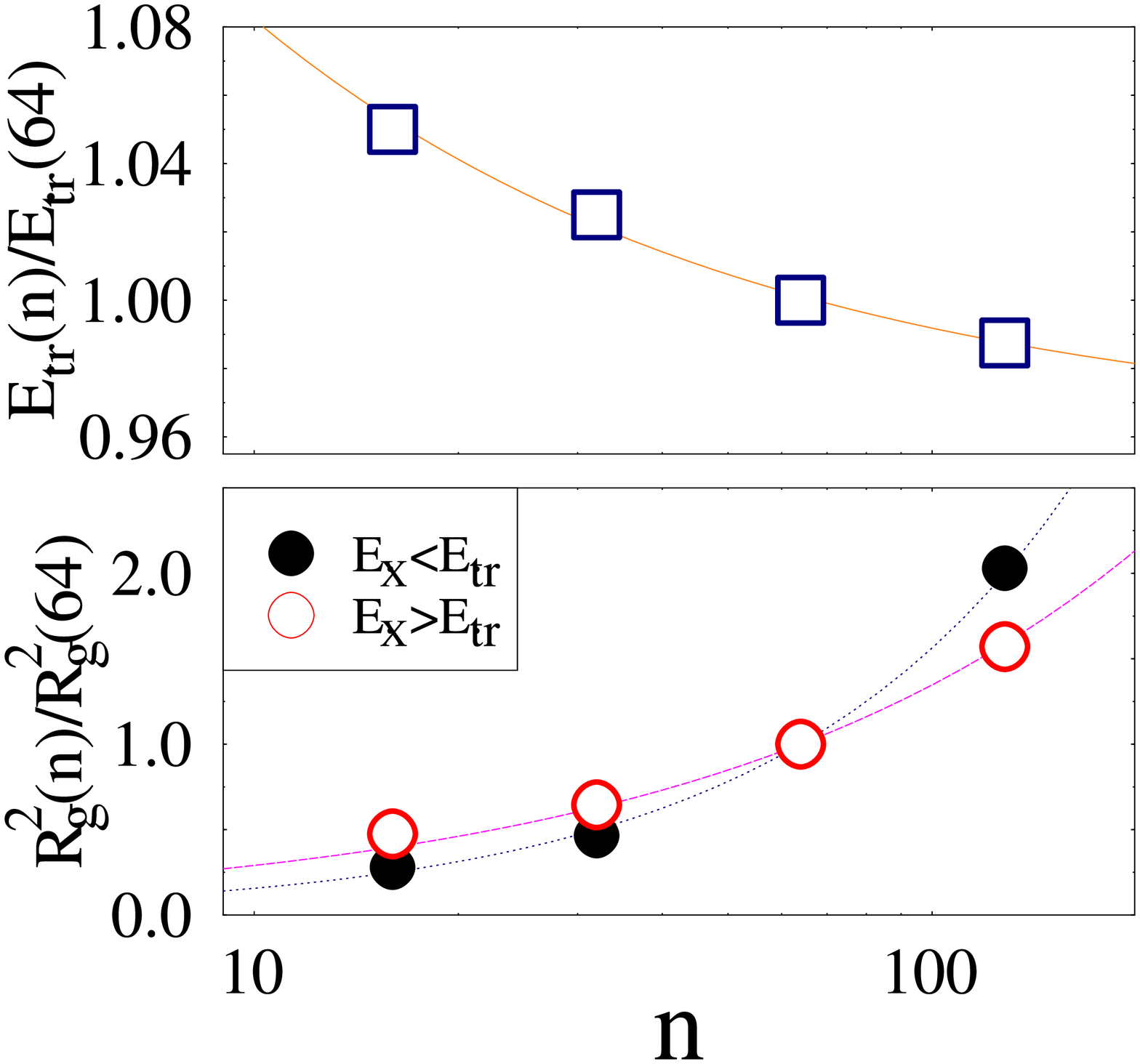}
      \caption{}
      \label{scaling}
   \end{center}
\end{figure}

We investigate, in particular, how the average gyration radius, $R_g$, and the threshold energy, $E_{tr}$, depend on $n$. For a matter of clarity, we refer to the case discussed in the left panel of Fig.\ref{equil_ex}, but similar features are found for the other cases presented in our paper. 
We, thus, consider a system with $c=0.04\%$ and $f=1/3$, and discuss first the case where 
$E_X=1kT$, i.e., the phase where the polymer is ``open" (see Fig.\ref{equil_ex} left panel). Under these circumstances, as shown in the lower panel of Fig.\ref{scaling}, $R_g$ scales with $n$ as a power law, $R_g \sim n^{\nu}$, with an exponent $\nu\sim 0.6$ which is in agreement with the random SAW scaling laws  \citep{EdwardsDoi,BinderHeermann}. Conversely, for $E_X=4kT$, i.e., in the ``looped" phase, $R_g$ scales as $n^{1/3}$ (see lower panel of Fig.\ref{scaling}), showing that the polymer is lumped in a compact conformation ($1/3$ is the inverse of the Euclidean dimension of the system). 
The threshold energy $E_{tr}$ has also a comparatively simple behavior with $n$ and appears to saturate to a finite value for large $n$. For instance, the threshold energy defined in the left panel of Fig.\ref{equil_ex} (where $E_{tr}(64)\simeq 3kT$) can be well fitted by a power law in $n$ (see upper panel of Fig.\ref{scaling}): $E_{tr}(n)=E_{tr}^{\infty}+A/n^B$, where $E_{tr}(n)$ is the value for a chain of size $n$, $E_{tr}^{\infty}$ the fitting value for an infinitely long chain, $A$ and $B$ a fitting coefficient and exponent (we find $E_{tr}^{\infty}\simeq 0.96E_{tr}(64)$, $A\simeq 0.47E_{tr}(64)$ and $B\sim 0.5$). 

Similar properties are found for the other quantities discussed in this paper. 
These checks outline the robustness of the picture discussed above and also support the idea that it is not an artifact of discretization, as a system in the continuum limit, i.e., on a finely divided lattice, 
should have an analogous behavior.

\subsection*{Interchromosomal segment interactions}

The mechanisms that drive other layers of spatial organization, including the colocalization of DNA sequences belonging to different chromosomes \citep{Pombo06} and the relative positioning of chromosomal territories \citep{Cremer2006,Lanctot2007,Misteli2007,Meaburn2007,Fraser2007,Gasser07}, can be shown to be very similar to those inducing stable loop formation within a single chromosome. Concentration/affinity acts in these cases as a ``thermodynamic switch" for segment colocalization and for chromosome positioning in a map. 

\begin{figure}
   \begin{center}
      \includegraphics*[width=5in]{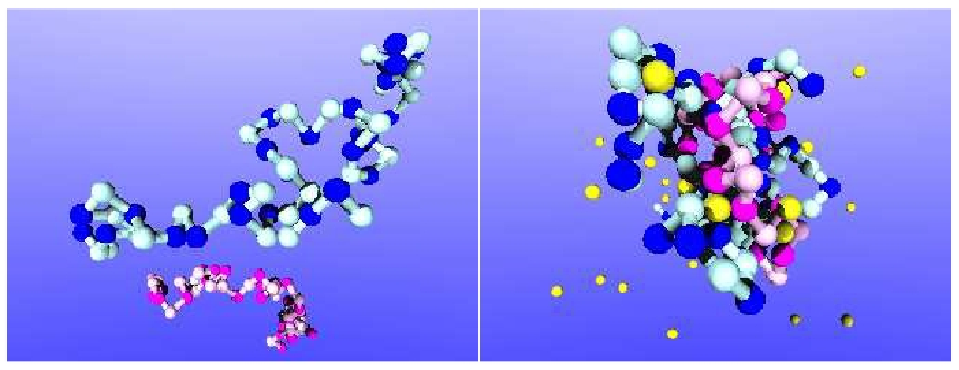}
      \caption{}
      \label{equil_picts2}
   \end{center}
\end{figure}

To such an aim, in an extension of the model described above, we now investigate the thermodynamic state of two SAW chains (representing either two distal sequences on the same chromosome or sequences on distinct chromosomes) with a fraction $f$ of binding sites (periodically placed) for a concentration, $c$, of molecules having an affinity, $E_X$, to both of them (see Fig.\ref{equil_picts2}); for simplicity, each molecule can bind once either polymer. 
The relative polymer positioning is given by their squared distance:
$d^2={1/(2n^2{\cal D})}\sum_{i,j=1}^{n} ({\bf r}^{(1)}_i-{\bf r}^{(2)}_j)^2$, 
where ${\bf r}^{(1)}_i$ (resp. ${\bf r}^{(2)}_i$) is the position of
bead $i$ in chromosome 1 (resp. 2), and ${\cal D}$ a normalization constant (here ${\cal D}$ is equal to the average square distance of two independent random SAW chains). The average value of $d^2$ is maximal when polymers float independently (i.e., $d^2=1$ in our normalization) and decreases drastically when all or parts of the chains become colocalized.

\begin{figure}
   \begin{center}
      \includegraphics*[width=2.25in,angle=-90]{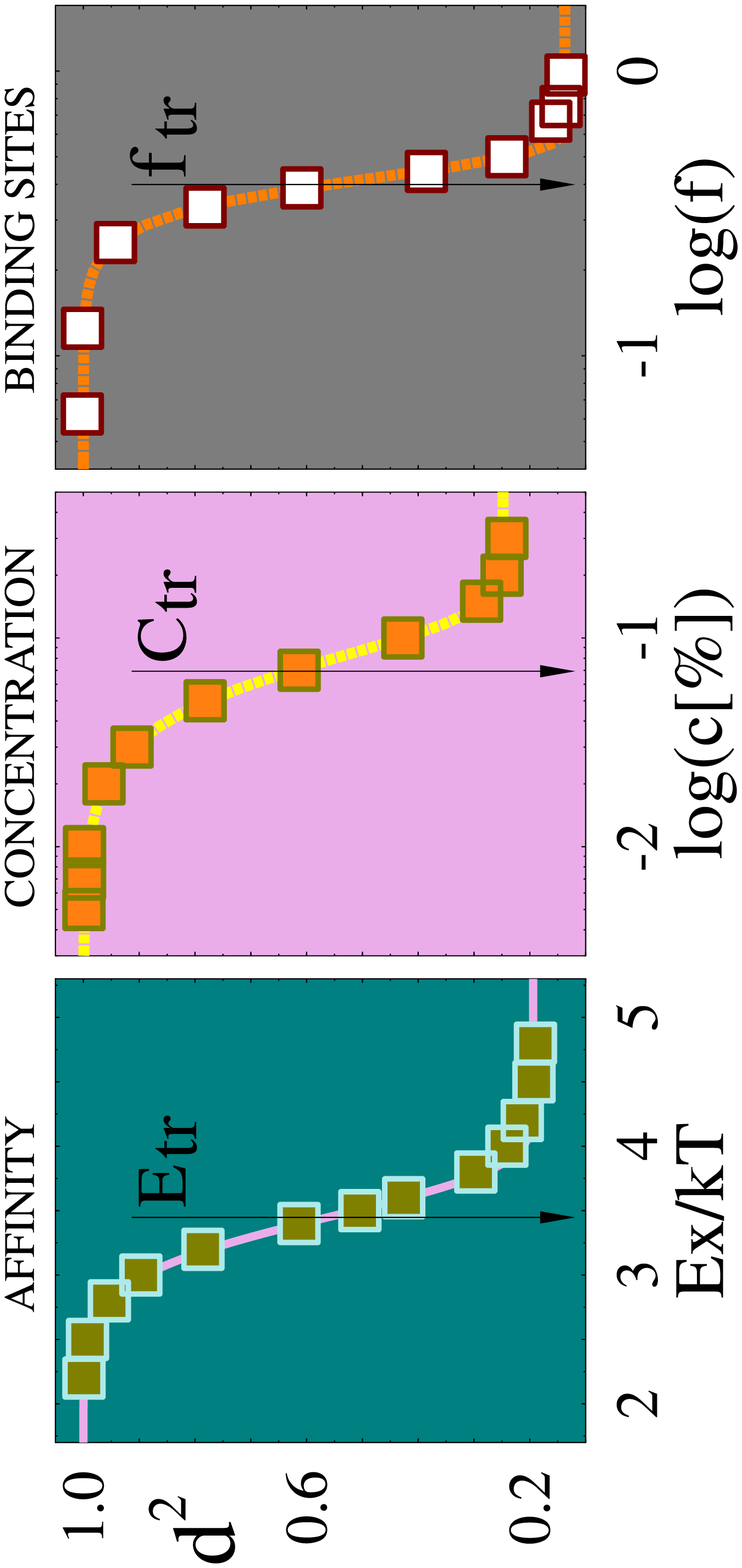}
      \caption{}
      \label{equil2_ex}
   \end{center}
\end{figure}

Regulation of $E_X$ can induce formation or release of stable physical contact between the polymers. Fig.\ref{equil2_ex} shows that when $E_X$ is below a threshold, $E_{tr}$, their equilibrium distance, $d^2$, has the same value found for two non-interacting Brownian SAW chains (i.e., $d^2=1$). 
This is the `random phase' where chromosomes move independently.
By thermodynamics mechanisms an effective attraction between the polymers is, instead, established when $E_X > E_{tr}$: physical contact is stable and $d^2$ drastically decreased, as the system enters the `colocalization phase'. 
The equilibrium distance is a function of $c$ as well (see Fig.\ref{equil2_ex}): when $c$ is below a threshold value, $c_{tr}$, a random distance is found between chromosomes (i.e., $d^2=1$). 
Colocalization is spontaneously attained, instead, when $c$ increases, as $d^2(c)$ approaches a plateau with a much smaller value. 
Finally, for a given $c$ and $E_X$, colocalization can be achieved only if the number of binding sites along the polymers is above a sharp threshold value, as shown in Fig.\ref{equil2_ex} where $d^2(f)$ is plotted. 

Alike loop architecture within a chromosome territory, the average distance of chromosome pairs can be controlled via thermodynamics mechanisms. The spatial association is attained when a phase transition line is crossed, corresponding to the point where entropy loss due to chain pairing is compensated by energy gain as both polymers are bound, the lower $E_X$ the higher the concentration, $c$, required. 

\subsection*{Assembly of chromosome territorial maps}

\begin{figure}
   \begin{center}
      \includegraphics*[width=3.25in]{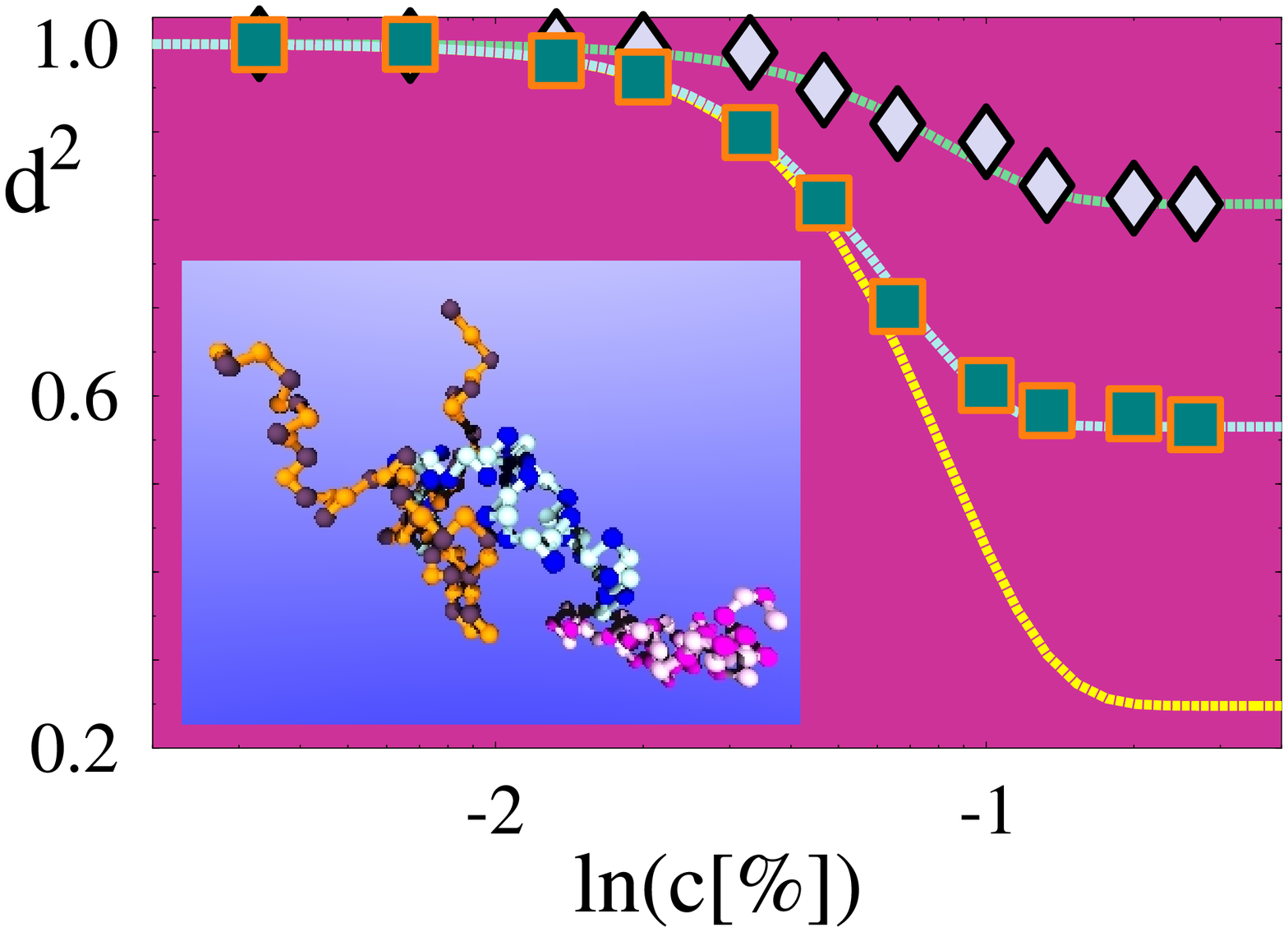}
      \caption{}
      \label{equil3_d}
   \end{center}
\end{figure}

Within the above picture, the relative positioning of chromosomal loci and territories can be understood by similar arguments. As an example, we considered (see inset in Fig.\ref{equil3_d}), the case with three SAW chains ($n=64$) having each a fraction, $f$, of binding sites ($f=1/2$, $E_X=4kT$): the sites on polymer 1 and 2 interact with a molecular factor (concentration $c_{12}$) which can bind once either chain; polymer 2 and 3 bind a different molecular factor of concentration $c_{23}$ (for definiteness, we only discuss the case where $c_{12}=c_{23}=c$). In order to illustrate the important effects of physical interference between chromosomes, in this model all molecular factors compete for the same sites on polymer 2. For the built in symmetry, polymer couples 1-2 and 2-3 behave similarly and have, on average, equal relative distances $d_{12}^2=d_{23}^2$ as a function of $c$ (see Fig.\ref{equil3_d}). Yet, since polymer 1 and 3 physically interfere when bridging with 2, in a competition for its binding sites, their distance is larger than the one found in the case with only two polymers under similar conditions (i.e., same $c$, $E_X$, $f$ and system size). The distance between 1 and 3, $d_{13}$, is in turn larger than $d_{12}=d_{23}$ because there is not a direct interaction. The three `chromosomes', thus, spontaneously find their position to form a (isosceles) `triangle' having sides of predefined length ($d_{12}$, $d_{23}$, $d_{13}$). 

Different patterns of relative positions can be attained by tuning the concentration/affinity switches, as the system architecture self-organize via thermodynamics pathways, funneling the interaction between sets of DNA binding sites and matching molecular mediators. When the number and length of chromosomal segments increase, the dynamics of the system to equilibrium can be slowed down by physical hindrance. This rises the speculation that the spatial organization of chromosomes in distinct territories and within territories (along with other mechanisms, e.g., the action of topoisomerases) may also serve the purpose of a faster and better control of their interaction and function, by reducing undesired entanglements.

\section*{Discussion}

Within the cell nucleus, in a striking example of self-organization, an astonishing number and diversity of DNA loci and molecular mediators are spatially orchestrated to form a complex and functional architecture involving regulatory cross talk between distant sites. We propose a simple conceptual framework, 
a ``thermodynamic switch model" of nuclear architecture, to understand some of its general features,  namely \citep{Meaburn2007}: 1) how a chromosome can fold up into a territory and how its looping is dynamically controlled by binding molecules; 2) how chromosomes interact and establish their relative positioning; 3) what are the regulatory principles and 4) the origin of the stochastic character of territorial maps.

Our model consists of a system of Self-Avoiding Walk polymers interacting with soluble molecular mediators. By use of Statistical Mechanics, 
we have shown that thermodynamics dictates pathways to complex pattern formation, via mechanisms such as ``thermodynamic switches" (see Fig.\ref{scheme}). This supports, on quantitative grounds, the idea that a variety of intra- and inter-chromosome interactions can be traced back to similar mechanisms. Looping and compaction, remote  sequence interactions and territorial segregated configurations correspond to thermodynamic states selected by appropriate values of concentrations/affinity of 
soluble mediators and by number and location of their attachment sites along chromosomes. After proper concentrations/affinities are set, the organization proceeds spontaneously with no energetic costs as the resources required, e.g., to rearrange even whole chromosomes, are provided by the surrounding thermal bath. 
Our picture explains, thus, how well described cell strategies of upregulation of DNA binding proteins or modification of chromatin structure can shape the genomic architecture and produce DNA colocalization and territories according to thermodynamically driven non random patterns.

Testable quantitative predictions are shown on the biological effects of alterations of genomic DNA sequences (such as deletions, insertions, chemical changes, etc.) and of their molecular mediators (concentrations, binding energies, etc.). 
In particular, the model highlights the fact that, at above threshold values of concentration, the interaction with low affinity molecular factors may be sufficient to drive the compaction of chromosomes into territories, and shows that the interaction of chromosomes with soluble mediators has the potential to impart a probabilistic relative arrangement to chromosomes. 
Our analysis reveals that molecular factors that act as bridges between two chromosomes may not only have the effect of pulling those close to each other, but may also displace non interacting chromosomes, so that these are farther away from each other than the ``random" distance. This result is thought-provoking in the light of experimental data \citep{Pombo06} showing that disruption of transcription can lead either to an increase or to a decrease of chromosome intermingling among specific couples of chromosomes, depending on what couple of chromosomes you look at. 
Allele-specific, parent-of-origin specific, and expression-specific DNA-DNA interactions have also been described \citep{Ling2006,Zhao2006,Maynard2008,Lomvardas2006,Takizawa2008}. 
In this context, our analysis could explain how imprinting and other allele-specific protein-DNA interactions may have the capacity to address homologous chromosomes to two different regions of the territory map.

A rough estimate of threshold molecular concentrations in real nuclei can be made 
from our predicted concentration values: here $c$ is the number of molecules per lattice site, so the number of molecules per unit volume is $c/d_0^3$, where $d_0$ is the linear lattice spacing constant. The molar concentration $\rho$ is obtained by dividing by the Avogadro number ${\cal N}_A$. Note that threshold concentrations depend on the binding energy $E_X$ (see, e.g., fig.~\ref{ph_diag}). For sake of definiteness, however, we can consider the case with $E_X\sim 2kT$ (see fig.~\ref{ph_diag}), where threshold concentrations are around $c=0.1\%\div 0.01\%$. Under the rough assumption that $d_0$ is a couple of orders of magnitude smaller than the nucleus diameter (i.e., $d_0\sim 10  nm$), a threshold molar concentration would be $\rho \sim 0.1\div 1\, \mu mole/litre$, which is consistent with typical experimental values of nuclear protein concentrations \citep{protein_crowding_1,protein_crowding_2}. Such estimation is very rough, but may help to further bridge this study with biological investigations.

Starting from experimental results showing that chromatin fiber at large genomic distances, above 1-2Mb, exhibits a leveling-off of the mean square distance between two DNA sites, 
a Gaussian ``Random loop" polymer model was recently proposed \citep{Bohn07}. 
To explain these observations, the model introduced the idea of long range interactions along the polymer, where a given number $P$ of couples of distant beads, randomly selected along the chain, are bound by an harmonic potential of amplitude $\kappa$.
The model investigated the mean distance between sites and the size of loops, and showed that the presence of random loops on all length scales explains the leveling-off of the mean square distance. 
That model is in close similarity with the present work where cross interactions of a fraction, $f$, of DNA sites are mediated by the binding of molecular factors and by the formation of bridges of energy $E_X$. In our case the number of interacting site couples also depends on the concentration of mediators, $c$. 
Interestingly, the case mainly investigated in \citep{Bohn07} has $\kappa/kT=1$, which is in the energy range we consider, although our site interaction is short ranged, while in \citep{Bohn07} it is an harmonic potential. 
Nonetheless, the number of interacting site in our model would correspond, in the notation of ref. \citep{Bohn07}, to a $P$ which is a (non trivial) function of $c$. These considerations can illustrate the agreement between the discovery in \citep{Bohn07} of the leveling-off of the mean square distance and our finding, for instance, that for $c$ above threshold, the polymer gyration radius doesn't attain the (self-avoiding) random walk value but saturates to much smaller values.

In real cells, passive and active regulatory mechanisms can cooperate, adding further layers of complexity \citep{Cremer2006,Lanctot2007,Misteli2007,Meaburn2007,Fraser2007,Gasser07}, while the list of molecules mediating chromatin organization is likely to include dedicated structural proteins, RNAs and, e.g., the transcription, replication, or repair machinery \citep{Cook2002,Osborne2004,Chakalova2005}. 
In our picture, specificity of interactions is obtained by specific molecular mediators binding to specific loci, while other general molecules could help the process. 
In the arrangement of specific binding sites along chromosomes and scaffolding elements, a variety of spatial patterns can be encoded \citep{CremerCremer} on an evolutionary time scale. Within a cell, patterns could be then dynamically selected by the combinatorial use of a set of mediators via the ineluctable, yet probabilistic, laws of thermodynamics \citep{MarenduzzoMicheletti}.

\bigskip

{\small Work supported by grant MIUR-FIRB RBNE01S29H, Network MRTN-CT-2003-504712. The authors declare no competing interests.}

\bibliography{nicodemi_bibtex1}

\clearpage
\section*{Figure Legends}

\subsubsection*{Figure~\ref{equil_picts}}
The figure shows two representative snapshots from our 3D computer simulations. In the {\bf left panel} a Self-Avoiding Walk (SAW) polymer is shown, as it floats randomly within the assigned volume without forming stable loops. In the {\bf right panel} the volume also contains a concentration $c=0.04\%$ of Brownian molecules (yellow) having an affinity $E_X=4kT$ for a fraction $f=1/3$ of the polymers beads (shown in a darker shade).  
As molecules can bind more than one polymer site, loops can be formed. However, they are stable, and confine the polymer in a closed territory (as in the case shown here), only if $c$ is above a threshold value (see Fig.\ref{equil_ex}). The SAW chains shown here comprise $n=64$ beads. 

\subsubsection*{Figure~\ref{equil_ex}}
The equilibrium average gyration radius, $R_g^2$, of the model polymer pictured in Fig.\ref{equil_picts}, depends on the affinity, $E_X$, of its binding sites for a set of molecular factors, on the concentration, $c$, of those factors, and on the fraction, $f$, of polymer beads which can bind molecules. $R_g$ represents the radius of a sphere enclosing the polymer: it has a maximum ($R_g^2=1$ in our normalization) when folding is random and a minimum when the polymer loops on itself in a lump (the horizontal red line is the radius of a compact sphere formed by the polymer). 
In the {\bf left panel}, $R_g^2$ is shown as a function of $E_X$, for a given value of $c$ and $f$ (here $c=0.04\%$, $f=1/3$). For $E_X$ below a threshold value, $E_{tr}\simeq 3kT$, $R_g^2$ is approximately $1$ and the polymer is on average open. For $E_X>E_{tr}$, $R_g^2$ collapses, as the polymer forms a looped territory. In the {\bf central panel}, $R_g^2$ is shown as a function of $c$, for a given $E_X$ and $f$ (here $E_X=4kT$,  $f=1/3$). Also in this case a threshold effects is observed ($c_{tr}\simeq 0.01\%$), although a broader crossover region exists where the level of folding can be tuned. The {\bf right panel} shows the sharp threshold of $R_g^2$ as a function of $f$ ($f_{tr}\simeq 0.1$, here $c=0.04\%$, $E_X=4kT$), illustrating that only in presence of multiple sites (i.e., above $f_{tr}$) the polymer can be folded in loops. 
In all the above cases, loops are thermodynamically stable only above the threshold values, as a consequence a  phase transition occurring in the system. By tuning affinities/concentrations, the cell can act, thus, on a ``thermodynamic switch" to form and release loops and territories. 

\subsubsection*{Figure~\ref{scaling}}
{\bf Lower panel: } The average gyration radius, $R_g$, relative to polymer model considered into the left panel of Fig.~\ref{equil_ex}, is plotted as a function of the polymer chain length $n$. The picture shows the ratio $R_g^2(n)/R_g^2(64)$ (since $n=64$ is the reference case dealt with in the rest of the paper) 
for $n=16,32,64,128$. 
In the phase where the polymer is ``open", i.e., for $E_X=1kT<E_{tr}$ (see left panel of  Fig.~\ref{equil_ex}), the average gyration radius, $R_g$ (filled circles), scales with $n$ as a power law $R_g \sim n^{\nu}$ with an exponent $\nu\sim 0.6$ \citep{EdwardsDoi,BinderHeermann} (superimposed fit, dashed line). 
In the ``looped" phase, i.e., for $E_X=4kT>E_{tr}$, $R_g$ (empty circles) scales as $n^{1/3}$ (superimposed fit, long dashed line), showing that the polymer is lumped in a compact conformation.
{\bf Upper panel: } The threshold energy, $E_{tr}$, relative to the left panel of Fig.~\ref{equil_ex}, is a function of the polymer chain length $n$. Here we plot the ratio  $E_{tr}(n)/E_{tr}(64)$ (where $E_{tr}(64)\simeq 3kT$). 
The superimposed fit is: $E_{tr}(n)=E_{tr}^{\infty}+A/n^B$, where $E_{tr}(n)$ is the threshold energy for a chain of size $n$, $E_{tr}^{\infty}\simeq 0.96E_{tr}(64)$ the extrapolated value for an infinitely long system, $A\simeq 0.47E_{tr}(64)$ and $B\sim 0.5$ a fitting coefficient and exponent.

\subsubsection*{Figure~\ref{ph_diag}}
The state of the polymer/chromosome (see Fig.\ref{equil_picts}) at thermodynamic equilibrium is summarized by this phase diagram in a range of values of `weak' biochemical affinities, $E_X$, and concentration, $c$, of its binding molecules (here $f=1/3$). When $E_X$ and $c$ are below the transition line, $E_{tr}(c)$ (empty circles), the polymer is `open' (as sketched in the inset) and no stable loops can be formed. Above threshold, instead, the system enters the region where the polymer is folded and `looped' on itself. 

\subsubsection*{Figure~\ref{equil_picts2}}
Two snapshots are shown from computer simulations of our two polymer model. In the {\bf left panel} the polymers float independently within the assigned volume. In the {\bf right panel} the volume also includes a concentration, $c=0.3\%$, of molecules (yellow particles) which can bind simultaneously each polymer once at any of their specific loci (darker sites, here in a fraction $f=1/2$ with affinity $E_X=4kT$). When $c$ is above a threshold value (see Fig.\ref{equil2_ex}), as in the case shown, thermodynamically stable bridges can be formed between the polymers, which spontaneously tend to pair parts of or all their chains. 

\subsubsection*{Figure~\ref{equil2_ex}}
The equilibrium average distance, $d^2$, of the two polymer model pictured in Fig.\ref{equil_picts2}, is a function of the affinity, $E_X$, of their binding sites for diffusing molecules, of the concentration, $c$, of molecules, and of the fraction, $f$, of polymer binding sites. 
In the {\bf left panel}, $d^2$ is plotted as a function of $E_X$ (here $c=0.3\%$, $f=1/2$). 
When $E_X$ is smaller than a threshold, $E_{tr}\simeq 3.5kT$, $d^2$ is maximal ($d^2=1$ in our normalization) and the polymers float independently one from the other. 
For $E_X>E_{tr}$, $d^2$ drastically decreases, as the polymers are spontaneously colocalized. 
In the {\bf central panel}, $d^2$ is shown as a function of $c$ (here $E_X=4kT$, $f=1/2$) and a threshold appears as well ($c_{tr}\simeq  0.07\%$), sourrounded by a crossover region. 
In the {\bf right panel}, the sharp threshold of $d^2$ as a function of $f$ is shown ($f_{tr}\simeq 0.4$, here $E_X=4kT$, $c=0.3\%$): only multiple binding sites, above $f_{tr}$, can achieve polymer colocalization. 
The mechanism driving polymer colocalization is an effective reciprocal attraction of thermodynamic origin, related to a phase transition: below threshold, molecules bridging by chance the polymers do not succeed in holding them in place; above threshold, bridges are thermodynamically stabilized. Molecular mediators act, then, as a ``thermodynamic switch" to spontaneous formation and release of polymer stable contacts. 

\subsubsection*{Figure~\ref{equil3_d}}
The relative positions of three polymers can be regulated by the concentration of specific molecular factors. 
{\bf Inset } A configuration is shown from our computer simulations of a three polymer model. A specific molecular factor can bind polymers 1 (pink) and 2 (blue), while a different factor binds polymers 2 and 3 (orange). Both molecular factors have here a concentration $c=0.13\%$ ($E_X=4kT$, $f=1/2$), but they are not shown for clarity. 
{\bf Main panel } The average distance between polymers 1-2, $d_{12}^2$ (squares), decreases as a function of $c$ (the distance between 2-3 equals $d_{12}^2$, and is not shown). As an indirect effect of the attraction within pairs 1-2 and 2-3, the distance between 1 and 3, $d_{13}^2$  (diamond), decreases as well, remaining, though, above $d_{12}^2$. The three polymers, thus, tend to form a triangle with two short equal edges (corresponding to $d_{12}$ and $d_{23}$) and a longer edge (i.e., $d_{13}$). In general, by tuning $c$, $E_X$ and $f$ a variety of configurational patterns can be spontaneously attained. 
Notably, since polymers 1 and 3 compete for bridging the sites of polymer 2, they physically interfere and $d_{12}^2$ is larger than in the case of an isolated couple (yellow lower line, from Fig.\ref{equil2_ex}). 
A proper spatial organization of chromosomes in territories and within territories could also help minimizing physical interference and entanglement. 

\begin{figure}
   \begin{center}
      \includegraphics*[width=3.25in]{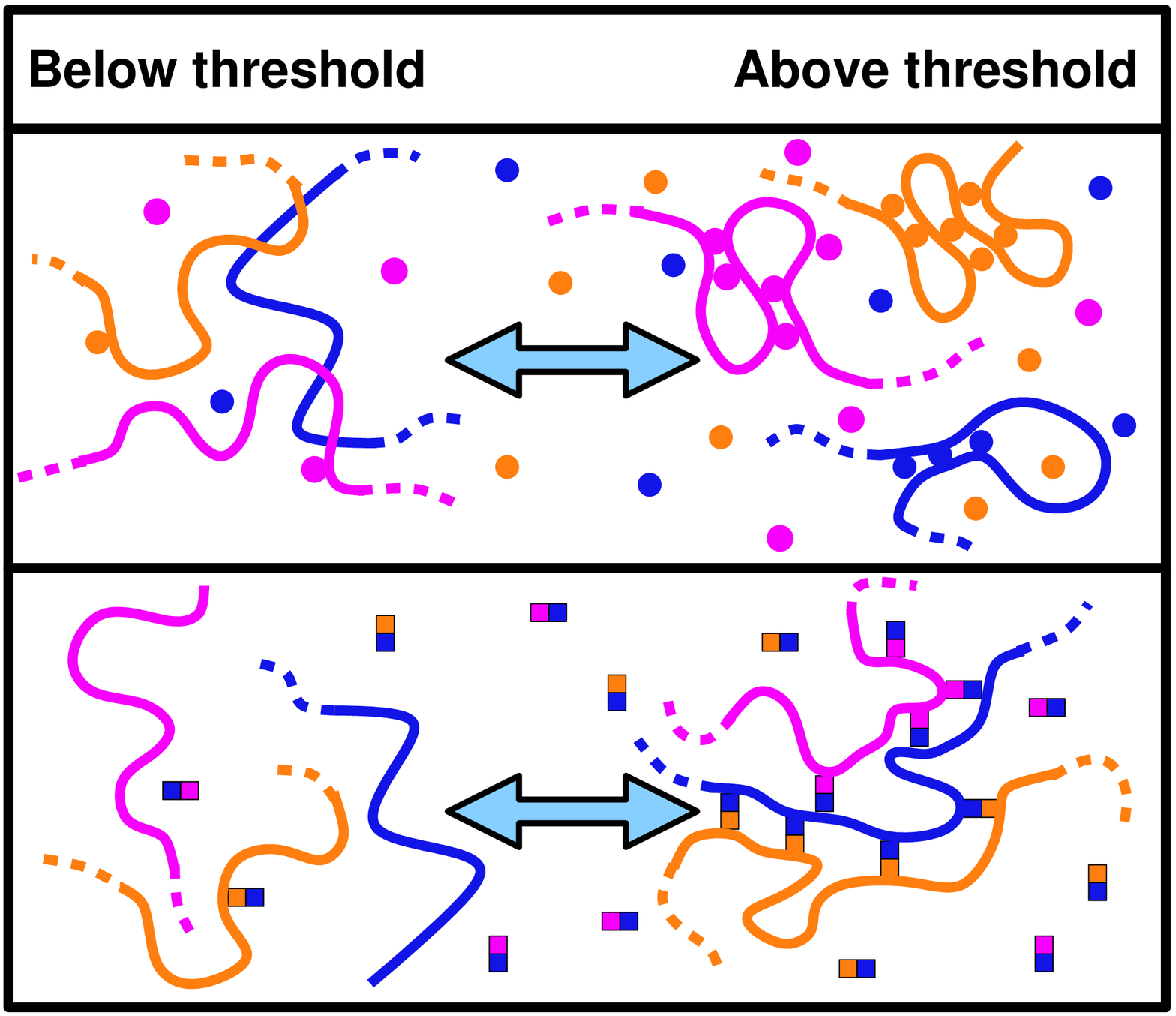}
      \caption{}
      \label{scheme}
   \end{center}
\end{figure}

\subsubsection*{Figure~\ref{scheme}}
Schematic illustration of ``thermodynamic switches" and their effects at different levels of system organization. {\bf Top panel} The assembly of chromosome loops is thermodynamically possible only when the concentration/affinity of binders (circles) exceeds precise threshold values. At that point, previously randomly and independently diffusing molecules and chromosomes spontaneously generate an organized pattern, in a process reversible by downregulation of the switch. Specific conformations can be attained by site specificity of a set of molecular mediators. 
{\bf Bottom panel} Similar threshold and self-organization mechanisms act for establishing contact between remote loci and, at a higher scale, relative positions of territories. A variety of patterns, encoded in the location of a number of binding sites along chromosomes, can be precisely selected via thermodynamics effects by a combinatorial use of a set of molecular mediators (rectangles).

\clearpage

\clearpage

\clearpage

\clearpage

\clearpage

\clearpage

\clearpage

\clearpage

\end{document}